\documentclass[11pt]{article}

\usepackage{amsmath, amssymb, amsthm}
\usepackage{graphicx}
\usepackage{hyperref}
\usepackage{geometry}
\usepackage{booktabs}
\usepackage{threeparttable}
\usepackage[numbers]{natbib}

\title{Dynamic Prediction of Joint Longitudinal-Survival Models Using a Similarity-Based Approach}
\author{
  Minzee Kim\thanks{Minzee Kim is the corresponding author and may be contacted at mz2kim@uwaterloo.ca.} \\[0.5ex]
  Department of Statistics \& Actuarial Science \\ 
  University of Waterloo, Waterloo, ON, Canada, N2L 3G1\\[2ex]
  Joel A.~Dubin \\[0.5ex]
  Department of Statistics \& Actuarial Science and \\ 
  School of Public Health Sciences\\
  University of Waterloo, Waterloo, ON, Canada, N2L 3G1
}
\date{} 

\begin{document}

\maketitle

\begin{abstract}
Longitudinal and time-to-event data are often analyzed in biomarker research to study the association between the longitudinal biomarker measurements and the event-time outcome, in which the longitudinal information contributes to the probability of the outcome of interest. An attractive nature of fitting a joint model on this type of data is that we can dynamically predict the survival probability as additional longitudinal information becomes available. We propose a new similarity-based method for the dynamic prediction of joint models where we consider training the model on only a targeted subset of the data to obtain an improved outcome prediction. Through comprehensive simulation study and an application to intensive care unit data, we demonstrate that the predictive performance of the dynamic prediction of joint models can be improved with our proposed similarity-based approach.
\end{abstract}


\maketitle

\renewcommand\thefootnote{}

\renewcommand\thefootnote{\fnsymbol{footnote}}
\setcounter{footnote}{1}

\section{Introduction}\label{introduction}

A biomarker is an indication and a measure of a medical state that can be measured objectively and reproducibly - it is ``any substance, structure, or process that can be measured in the body or its products and influence or predict the incidence of response or disease'' \cite{worldinternational}. In particular, biomarkers are used in research for disease diagnosis and personalized medication and treatment \cite{gosho2012study}. Longitudinal and time-to-event data are often analyzed in biomarker research to study the association between the longitudinal biomarker measurements and an event-time response, in which the longitudinal information contributes to the probability of the outcome of interest \cite{rizopoulos2012joint}. More specifically, in follow-up studies, different measurements are collected for each individual over time, including longitudinal responses and the time until an event occurs for that individual. It may be of interest to study the association between the longitudinal biomarker information and time-to-event response, instead of conducting separate analyses for each, which is often the goal of biomarker research where studies aim to identify biomarkers with strong prognostic capabilities for an event-time response \cite{rizopoulos2012joint}. For example, T4 and TSH (thyroid-stimulating hormone) levels are measured from blood samples to monitor patients at risk of thyroid cancer and to track the disease progression by investigating the association between these biomarkers and the time of cancer development. Another unique characteristic of follow-up studies is that it has a dynamic nature - the rate of progression is different between each individual and it also dynamically changes over time uniquely for individuals \cite{rizopoulos2012joint}. 

Joint longitudinal-survival (JLS) models were developed to accommodate the unique characteristics of longitudinal and time-to-event data. They are used to efficiently study the effects of endogenous time-varying covariates on the survival response; we can study the strength of association between the hazard of the event and the time-varying covariate by implementing joint models. Furthermore, joint models can dynamically predict the survival probability as additional longitudinal measurements become available, which reflects the unique nature of follow-up studies \cite{JSSv072i07}. That is, as time progresses and additional information becomes available for each participant, we can update the survival prediction for each participant using the new information. There has been much development in JLS modeling in the past 35 years; numerous extensions have been implemented, including flexible modeling of longitudinal trajectories, incorporating multiple longitudinal markers, and modeling multiple failure times \cite{rizopoulos2012joint}. In this work, we propose a new method as an extension of JLS modeling by using a similarity-based approach to improve the dynamic predictive performance of JLS models. 

Patient similarity has been considered a groundbreaking new paradigm and has been implemented to create customized models that lead to better prediction performance, particularly with model discrimination \cite{lee2015personalized}. In recent years, there has been much interest in applying patient similarity to biomedicine, where a sharp increase in electronic health records (EHR) has resulted in a sizeable amount of available patient information \cite{sharafoddini2017patient}. Often, we use these data to fit a model for patient response prediction as it is directly related to intervention and treatment selection, care planning, and resource allocation. Traditionally, large and heterogeneous data such as EHR were analyzed for predictions. However, using the entire training data shows the ``average best choice''; thus, physicians and practitioners cannot rely on just this information when they want to make a prediction for a new patient with characteristics that diverge from the average. This is the motivation behind Personalized Predictive Modeling (PPM) based on patient similarity - the goal is to provide personalized predictions by identifying patients in training data similar to an index patient for whom we are trying to make a prediction \cite{sharafoddini2017patient}. Past research has shown that implementing patient similarity can improve the predictive accuracy of different models, such as logistic regression, decision trees, and random survival forests \cite{lee2015personalized, xu2019similarity}. 

Quantifying similarity between the patients is an important step in PPM to improve the predictive accuracy of the model. Numerous methods were developed to measure similarity, including but not limited to distance-based similarity metrics such as the Euclidean distance and Mahalanobis distance; cosine similarity metrics; correlation-based similarity metrics where the correlation coefficient was used to identify similar patients; and cluster-based algorithms that group the patients in the training data based on their profiles \cite{sharafoddini2017patient}. In this work, we implement the cosine similarity metric (CSM) which measures the cosine of the angle between the predictor vectors of two participants. More specifically, it is expressed as
\begin{equation*} 
CSM(S_i, S_j) = \frac{\mathbf{S_i} \cdot \mathbf{S_j}}{\|\mathbf{S_i}\| \|\mathbf{S_j}\|},
\end{equation*}
where $\mathbf{S_i}$ and $\mathbf{S_j}$ are the predictor vectors of Participant $i$ and Participant $j$, respectively, and $|| \cdot ||$ represents the Euclidean vector magnitude. Since the patient similarity is quantified as an angle between two vectors, it is normalized to be between -1, representing minimum similarity where the two vectors would be in exactly opposite directions, and 1, representing the maximum similarity where the two vectors are identical \cite{lee2015personalized}. One convenient feature of using cosine similarity over Euclidean or Mahalanobis distance is that it can naturally accommodate categorical variables. By calculating the CSM of an index participant with each of the participants in a training data set, we can rank the participants in the training data set in descending order to identify the participants that are the most similar to the index participant. 

By training the model on a more personalized subset of data consisting of similar individuals, we can increase the predictive accuracy; however, we must strike a balance in choosing the right sample size. We may offset the positive results from the patient similarity and develop a model with worse prediction accuracy than the original model when we increase personalization of the data and limit the sample size to be too small \cite{lee2015personalized}. Krikella and Dubin empirically showed that though model discrimination tends to improve as the size of the subpopulation decreases, model calibration may be more complicated and settling on a size of the subpopulation to optimize calibration is less obvious \cite{krikella2025personalized}. This said, sometimes we consider the Brier score to evaluate predictive model performance, which is a summary measure that comprises both discrimination and calibration. In this work, we aim to show that the prediction of joint models for longitudinal and survival data measured by the Brier score (i.e., specifically, time-varying Brier score \cite{henderson2002identification}), can be improved using a similarity-based approach. We use cross-validation to determine the optimal subpopulation size that produces the best prediction accuracy in the joint model. We also demonstrate that the Brier score of the joint model can be decreased when we train the model on the subpopulation of the most similar individuals in the training data to each of the individuals in the testing data. Thus, we make a novel contribution to the literature by improving the predictive performance of joint models using PPM and evaluating model performance across varying training data sizes consisting of similar observations through cross-validation.

This paper is organized as follows. Section \ref{Methods} begins with a brief discussion on fitting a joint model, evaluating its predictive accuracy, and handling time-varying predictors when calculating the CSM. These components are used to implement our proposed algorithm for improving the predictive accuracy of the joint model using a similarity-based approach, presented in Section \ref{sec:algorithm}. We then describe the details of a comprehensive simulation study in Section \ref{sec:simulation}, including data generation procedure and our simulation results. In Section \ref{sec:dataanalysis}, we apply our proposed algorithm in Section \ref{sec:algorithm} to a dataset obtained from the publicly-available eICU Collaborative Research Database \cite{pollard2018eicu} to attempt to confirm the results of our simulation study. Finally, we conclude with a discussion of our findings and future work in Section \ref{sec:discussion}.

\section{Methods}\label{Methods}
\subsection{Joint Models for Longitudinal and Time-to-event Data} \label{sec:JM}

Joint models for longitudinal and time-to-event data can accommodate endogenous covariates in survival analysis. We can use longitudinal information to better aid with survival prediction, over just using baseline data, and can dynamically update the survival prediction as more longitudinal data becomes available. A standard relative risk model to jointly fit longitudinal and time-to-event data is defined as the following \citep{rizopoulosJMbayes2}:
\begin{equation} \label{jm}
h_i(t | \textbf{M}_i(t)) = h_0(t) \exp\left\{\boldsymbol{\gamma} ^T \mathbf{w}_i + \alpha \textbf{m}_i(t)  \right\},
\end{equation}
where $\textbf{M}_i(t)$ is the longitudinal history of individual $i$ such that $\textbf{M}_i(t) = \{ \textbf{m}_i(s), 0 \leq s < t \}$, $\alpha$ quantifies the association between the endogenous covariate and the risk of an event, $h_0(t)$ is the baseline hazard function, and $\mathbf{w}_i$ is a vector of baseline covariates with corresponding regression parameters $\boldsymbol{\gamma}$ in the Cox proportional hazards model \citep{cox1972regression}. $\textbf{m}_i(t)$ is defined from a linear mixed effects model, where 
\begin{align*} 
\mathbf{Y}_i(t) &= \textbf{m}_i(t) + \boldsymbol{\epsilon}_i(t) \nonumber\\
    &= \mathbf{X}^T_i(t) \boldsymbol{\beta} + \mathbf{Z}_i^T(t) \mathbf{b}_i + \boldsymbol{\epsilon}_i(t).
\end{align*}

\noindent Here, $\mathbf{X}_{i}^T(t)$ is  a $(n_i \times p)$ matrix of covariates; $\boldsymbol{\beta}$ is a $(p \times 1)$ vector of fixed effects; $\mathbf{Z}_{i}^T(t)$ is a $(n_i \times q)$ matrix of covariates, typically a subset of $\mathbf{X}_i^T(t)$, with $q \leq p$; $\mathbf{b}_i$ is a $(q \times 1)$ vector of random effects with $\mathbf{b}_i \sim MVN(\mathbf{0}, \mathbf{D})$; and $\boldsymbol{\epsilon}_{i} $ is the random error associated with individual $i$ such that $\boldsymbol{\epsilon}_{i} \sim MVN(\mathbf{0}, \sigma^2 \mathbf{I}_{n_i})$ \citep{fitzmaurice2012applied}. This last assumption on the model error terms, referred to as the ``conditional independence" assumption, can be relaxed, if the model so warrants, by imposing a (serial) correlation structure on these error terms. The \textit{JMbayes2} package in R is often used to fit joint models \cite{RJMbayes2}.

Then the joint distribution of the response, survival time, and the censoring factor is defined as the following form, 
\begin{equation} \label{jointdist}
    p(\mathbf{Y}_i, T_i, \delta_i) = \int p(\mathbf{Y}_i | \mathbf{b}_i) \left\{ h(T_i|\mathbf{b}_i)^{\delta_i} S(T_i|\mathbf{b}_i) \right \} p(\mathbf{b}_i) d \mathbf{b}_i,
\end{equation}
with $p(\cdot)$ and $S(\cdot)$ being the density function and survival function, respectively \citep{tsiatis2004joint}. Note that in the density function $p(\mathbf{Y}_i | \mathbf{b}_i)$, the random effects $\mathbf{b}_i$ explain all the interdependencies; conditional on $\mathbf{b}_i$, the longitudinal measurement is independent of the time-to-event response, and the repeated measurements of the longitudinal data are independent of each other \citep{rizopoulosJMbayes2}. 

Unlike the standard Cox model, however, joint models require a full specification of the joint distribution, meaning that we need to make certain assumptions for the baseline hazard. General advice by Rizopoulos \citep{rizopoulosJMbayes2} is to use a flexible parametric model for the baseline hazard such that 
\begin{equation}\label{paramh0t}
\log h_0(t) = \boldsymbol{\gamma}_{h_0, 0} + \sum^Q_{q=1} \boldsymbol{\gamma}_{h_0, q} B_q(t, v),
\end{equation}
where $B_q(t, v)$ is the $q$-th basis function of a B-spline with knots $v_1, ..., v_Q$ and $ \boldsymbol{\gamma}_{h_0}$ is a vector of spline coefficients. We can increase the flexibility of the model by increasing the number of knots $Q$ - however, we prefer to balance bias and variance and avoid overfitting \citep{JSSv072i07}.

There have been many approaches for estimating the model parameters, including both maximum likelihood and Bayesian. In this work, we utilize the Bayesian method \citep{rizopoulosJMbayes2, mauff2020joint, rizopoulos2016personalized}. We assume that conditional on the random effects, the longitudinal measurements are independent of the event times and the longitudinal measurements are independent of each other. Under the Bayesian paradigm, both $\boldsymbol{\theta}$, which is the vector of parameters in the joint model including $\boldsymbol{\gamma}, \alpha, \boldsymbol{\beta}$, the variance and covariance parameters of $\mathbf{D}$, and $\sigma$, and $\{\mathbf{b}_i, i = 1,..., n \}$ are regarded as parameters. Then, inference is based on the following full posterior distribution: 
\begin{align} \label{Bayesian param}
    p(\boldsymbol{\theta}, \mathbf{b} | \mathbf{T}, \boldsymbol{\delta}, \mathbf{Y}) &= \frac{\prod_i p(T_i, \delta_i|\mathbf{b}_i;\boldsymbol{\theta}) p(\mathbf{Y}_i | \mathbf{b}_i;\boldsymbol{\theta}) p(\mathbf{b}_i;\boldsymbol{\theta}) p(\boldsymbol{\theta}) }{\prod_i p(T_i, \delta_i, \mathbf{Y}_i)} \nonumber\\
    &\propto \prod^n_{i=1} \left \{ p(T_i, \delta_i | \mathbf{b}_i;\boldsymbol{\theta}) p(\mathbf{Y}_i |\mathbf{b}_i;\boldsymbol{\theta}) p(\mathbf{b}_i;\boldsymbol{\theta}) \right \} p(\boldsymbol{\theta}).
\end{align}

The \textit{JMbayes2} package in R by Rizopoulos et al. \citep{rizopoulosJMbayes2} is often used to fit joint models. As there is no closed-form solution for integrating Equation (\ref{Bayesian param}) over the random effects $\textbf{b}_i$, \textit{JMbayes2} \citep{RJMbayes2} package uses Gibbs and Metropolis-Hastings MCMC algorithm with Robbins-Monro adaptive optimal scaling to estimate the parameters in $\boldsymbol{\theta}$. 

\subsection{Predictive Accuracy of Joint Models} \label{sec:Prediction}

Often, the motivation behind fitting a statistical model to a dataset is to predict the response. With longitudinal and time-to-event data, it offers a unique advantage that we can dynamically predict the probability of an event; thus, we can use all the available information at hand, including both the baseline information and accumulated biomarker measurements, to update our predictions of the probability of the event as more biomarker measurements become available. Furthermore, we can make a personalized prediction that is specific to the individual of interest. With growing interest in prognostic models and personalized medicine in medical care, dynamic prediction of joint models has been especially helpful in informing physicians about future prospects of a patient to adjust medical care \citep{barrett2017dynamic}.

Formally put, we have longitudinal measurements available up to time point $t$ for a new participant $j$,
\begin{equation*}
\mathbf{Y}_j(t) = \{y_j(s), 0 \leq s \leq t \},
\end{equation*}
and we are interested in the survival probability of patient $j$ at some future time $u$ using both baseline covariates and longitudinal information available until time $t$, which can be written as
\begin{equation*} 
\pi_j(u|t) = P \{T_j^* \geq u | T^*_j > t, \mathbf{Y}_j(t), D_n \},
\end{equation*}
where $T^*_j$ is the true underlying event time of individual $j$, $u > t$, and $D_n$ is the sample in which the joint model was fitted.

In general, calibration and discrimination are used to evaluate the performance of the model, where the former measures how well the model predicts the observed data and how closely the predicted probability agrees numerically with the true response \citep{d2003evaluation} and the latter focuses on how well the model can discriminate between participants who experienced the event and those who did not \citep{d2003evaluation}. Brier scores are also often used to evaluate model performance, as they can be decomposed into separate components that respectively capture both discrimination and calibration \citep{krikella2025personalized, murphy1973new}. Two things must be considered when measuring the predictive accuracy of joint models for longitudinal and time-to-event data. The first is censoring - censoring may depend on the observed longitudinal measurements or on the baseline covariates, so the performance measures must be corrected for censoring using model-based estimators of the censoring distribution \citep{rizopoulos2017dynamic}. The second is the fact that due to its dynamic nature, the model predictions are updated as more longitudinal information becomes available. Then, we redefine discrimination so that we are interested in the event occurrence in a time frame $(t, t+\Delta t]$ when we have longitudinal measurements for participant $j$ until time $t$. For calibration and Brier score, we are interested in predicting the probability of an event at time $t+ \Delta t$ given longitudinal information until $t$.  

In this work, we have decided to use the time-dependent Brier score to measure the predictive performance since, as just conveyed above, it captures both discrimination and calibration, though we will discuss utilizing these performance measures separately in the Discussion section \citep{krikella2025personalized, murphy1973new}. The expected quadratic error of prediction, or the Brier score, conditional on time $u > t$, is
\begin{equation*}
PE(u|t) = E \left[\{ N_j(u) - \pi_j(u | t) \}^2 \right],
\end{equation*}
where $N_j(t) = I(T_j^* > t)$, or the event status at time $t$, and $u = t + \Delta t$.  Then the estimate of the Brier score that accounts for censoring is
\begin{equation} \label{bs}
\widehat{PE}(u | t) = \left( R(t) \right)^{-1} \sum_{j: T_j \geq t} (\widehat{L}(u | t)),
\end{equation}
where 
\begin{align} \label{loss}
\widehat{L}(u | t) &= 
 I(T_j \geq u) \left(1 - \hat{\pi}_j (u|t)\right)^2 \nonumber\\
 & + \delta_j I(T_j < u) \left(0 - \hat{\pi}_j (u|t)\right)^2 \nonumber\\
 & + (1-\delta_j) I(T_j < u) \left\{  \hat{\pi}_j(u|T_j) \left(1 - \hat{\pi}_j (u|t)\right)^2 + \left(1 - \hat{\pi}_j (u|T_j)\right) \left(0 - \hat{\pi}_j (u|t)\right)^2        \right\} ,
\end{align}
\noindent
and $R(t)$ is the number of participants at risk at time $t$ \citep{henderson2002identification}. In Equation \eqref{loss} combined with Equation \eqref{bs}, the first and second terms of $\widehat{L}(u | t)$ correspond to patients who were alive after time $u$ and dead in the interval $[t, u)$, respectively, and the third term corresponds to those who were censored in the interval $[t, u)$ \citep{rizopoulos2017dynamic}.

\subsection{Calculating CSM with Time-Varying Covariates} \label{sec:PPM}

Recall that CSM is expressed as the following, which defines the similarity between the index participant and the participant in the training data:

\begin{equation*} \label{csm}
CSM(S_i, S_j) = \frac{\mathbf{S_i} \cdot \mathbf{S_j}}{\|\mathbf{S_i}\| \|\mathbf{S_j}\|}.
\end{equation*}
\noindent
We need to identify $\mathbf{S_i}$ and $\mathbf{S_j}$, the predictor vectors for Participant $i$ and Participant $j$. While it is straightforward to include the baseline covariates as part of the vectors, it can be advantageous to take into account longitudinal measurements as well. To represent the multiple longitudinal measurements into a finite number of vectors, we use functional principal component analysis (FPCA), a multivariate analysis technique that extracts information from functional data, such as longitudinal data \citep{wang2016functional}. It reduces the dimension of data with a large number of interrelated variables and transforms the data to a new set of variables, called functional principal components, while retaining as much of the total variation as possible \citep{ullah2013applications}. The principal components are uncorrelated and can be ordered so that the first few will retain most of the variation present in the original variable. This is implemented using the Principal Analysis by Conditional Estimation (PACE) algorithm in the R package \textit{fdapace} \citep{fdapace} where the expansion for the longitudinal measurement $Y_i(t)$ with only the first $r$ eigenfunctions ($\hat{\phi}_k(t), k = 1, 2, ..., r)$ is 
\begin{equation*}
\hat{Y}_i^{r}(t) = \hat{\mu}(t) + \sum^{r}_{k=1} \hat{\xi}_{ik} \hat{\phi}_k(t) ,   
\end{equation*}
where $\hat{\mu}(t)$ is the estimated mean function and $\hat{\xi}_{ik}$ are the FPCs. Therefore, in our simulation to be discussed in Section \ref{sec:simulation}, we used FPCA to reduce the dimensionality of the longitudinal data and use FPCs to explain the variation in the trajectories, which are used in the calculation of the cosine similarity metric along with the baseline covariates to quantify the similarity between the participants in the study. $r$ can be chosen using a criterion such as the number of FPCs that exceed a certain threshold of explained variation (e.g., 95\%) in a given longitudinal variable. Note that with the PACE algorithm, we can apply FPCA to irregular and even sparse longitudinal measurements. Even when measurements are missing at certain time points for a participant, PACE borrows strength from the entire data to estimate the global mean and covariance, and uses these estimates to calculate the FPCs by conditioning on the observed data. 

\subsection{Algorithm} \label{sec:algorithm}

In this section, we describe our new proposed algorithm to improve the dynamic prediction of joint models for longitudinal and time-to-event data at a specific time point $t$ for a specific future time $u$ that has some subject-matter interest, called \textit{similarity-based dynamic prediction using joint models}.

\begin{enumerate}
    \item Let $D_n$ be the training data that contains longitudinal and survival information for $n$ individuals. Then, for each individual $j$ in the testing data, we use the cosine similarity metric and the features of the individuals in $D_n$ to calculate the similarity between the participant in $D_n$ with individual $j$.
    \item Let $M_p \in (0, 1]$ be the proportion of the training data for fitting a personalized model. Let $m = M_p \times n$ be the subpopulation size for fitting the personalized model. We define $D^*_{m,j}$, which is the subset of the training data containing $m$ individuals that are the most similar to patient $j$ in the testing data, determined by the cosine similarity metric. 
    \item Fit a joint model with $D^*_{m,j}$ using the following equation:
    \begin{equation}\label{jointmodel*}
        h^*_{i}(t | \textbf{M}^*_{i}(t)) = h^*_0(t) \exp\left\{\boldsymbol{\gamma} ^T \mathbf{w}^*_i + \alpha \textbf{m}^*_i(t)  \right\},
    \end{equation} where $ \textbf{M}^*_{i}(t)$ is the longitudinal history of individual $i$ in $D^*_{m,j}$, $\mathbf{w}^*_i$ is the baseline covariate vector of individual $i$ in $D^*_{m,j}$, and $\textbf{m}_{i*}(t) = \mathbf{X}^{T}_{i*}(t) \boldsymbol{\beta}^* + \mathbf{Z}^{*T}_i(t) \mathbf{b}_{i*}$.
    \item  Using the fitted joint model in Step 3, calculate the probability of survival at time $u$ given biomarker measurements until time $t$ with $D^*_{m,j}$:
    \begin{equation} \label{bayesianpiut*}
    \pi^*_j(u|t) = P\{T^*_j \geq u | T^*_j \geq t, \mathbf{Y}^*_j(t), D^*_{m,j}\}.
    \end{equation}
\end{enumerate}

\noindent
In other words, by training the joint model on $D^*_{m,j}$ instead of $D_n$, we are using personalized data that contains $M_p \times 100\%$ of the original data $D_n$ that is the most similar to the testing participant $j$ to dynamically predict its response.

By decreasing $M_p$, we can create personalized training data for predicting the response of participant $j$ as it will contain smaller and smaller subsets of participants from $D_n$ that are more similar to participant $j$. However, a smaller dataset may cause other problems, such as overfitting the data (e.g., if too many covariates are considered relative to sample size), having optimization problems for fitting the underlying LME model, or fitting an inadequate model due to a small sample size. Therefore, it is of interest to identify the optimal $M_p$ that will maximize the accuracy in predicting the survival probability at future time $u$ given baseline covariates and longitudinal information available until time $t$. Below, we present the algorithm to tune for $M_p$ using R-repeated K-fold cross-validation:

\begin{enumerate}
    
     \item Split the data into $K$ different folds for $K$-fold cross-validation.
     
     \item Let $k$ be the hold-out fold for testing data, and the other $K-1$ folds be the training data.
     
     \item For training and testing data separately, apply FPCA using the PACE algorithm to extract the first few FPCs for the longitudinal measurement that cumulatively explain at least $S$\% of the total variation in the trajectory of the longitudinal marker. We use $S=95\%$ as the total variation of the longitudinal biomarker to be used by the FPCs in our simulation study below. Note that we use the longitudinal measurements available up to time $t$ to calculate the first few FPCs to avoid data leakage.

     \item Consider a grid of $M_p$ values to tune $M_p$.  

     \begin{enumerate}
         \item For an index participant $j$ in the testing data, calculate $CSM(S_j, S_i)$ for all participants $i$ in the training data. For the features to calculate the similarity between two participants using CSM, we consider the baseline covariates and FPCs to explain the variation in the trajectory of the time-varying covariates calculated in Step 2.
         
         \item Sort the participants in the training data in descending order by the similarity metric.

         \item Fit the joint model in Equation \eqref{jointmodel*} with $D^*_{m,j}$, or the top $M_p \times 100\%$ of the sorted training data. For a pre-specified $t$ and $u > t$, calculate the loss function for participant $j$ defined as the following:
         \begin{align*} \label{loss*}
\widehat{L}^*(u | t) &= 
 I(T_j \geq u) \left(1 - \hat{\pi}^*_j (u|t)\right)^2 \nonumber\\
 & + \delta_j I(T_j < u) \left(0 - \hat{\pi}^*_j (u|t)\right)^2 \nonumber\\
 & + (1-\delta_j) I(T_j < u) \left\{  \hat{\pi}^*_j(u|T_j) \left(1 - \hat{\pi}^*_j (u|t)\right)^2 + \left(1 - \hat{\pi}^*_j (u|T_j)\right) \left(0 - \hat{\pi}^*_j (u|t)\right)^2        \right\} ,
\end{align*} where $\hat{\pi}^*_j (u|t)$ is defined in Equation \eqref{bayesianpiut*}.

         \item Repeat Step 4(a) to Step 4(c) for every participant $j$ in the testing data and average over the values calculated in Step 4(c) to find the Brier score given by the following equation:
         \begin{equation*} \label{bs*}
\widehat{PE}^*(u | t) = \left( R(t) \right)^{-1} \sum_{j: T_j \geq t} (\widehat{L}^*(u | t)).
\end{equation*}
         
     \end{enumerate}

     \item Repeat Steps 3-4 for each $k$ in the $K$ folds. Average the value of the Brier scores over the $K$ folds.

     \item Repeat Step 2 to Step 5 $R$ times and average over the repeated $K$-fold cross validation runs. The optimal $M_p$ is the value with the smallest estimated Brier score for calculating the dynamic prediction at a specific time point $t$ for a specific future time $u$ that has some subject-matter interest.

    \item In the case where there is a desire to evaluate prediction performance on a hold-out validation sample, use the $M_p$ that was tuned using the training-testing data outlined in the steps above. We consider this step in the data analysis section in Section \ref{sec:dataanalysis}.
\end{enumerate}

\section{Simulations}\label{sec:simulation}

Below, we describe the simulation study performed to show how we can improve the dynamic predictive performance of joint models using a similarity-based approach. Tuning for the optimal subpopulation size using 5-fold cross-validation repeated ten times for each dataset under dynamic prediction for joint modeling is very computationally extensive. Therefore, we generate five datasets for two different simulation scenarios in this simulation study. Still, we plan to increase the number of repeats of the cross-validation and the simulation scenarios with more efficient coding in future research. We consider five possible values for the proportion of the training data to determine the size of the targeted subpopulation data - $20\%, 40\%, 60\%, 80\%,$ and $100\%$ - and calculated the Brier score of the model for predicting the response at time $u > t$ with biomarker measurements available until time $t = 1$. While it is useful to use longitudinal information to update the survival prediction, we cannot always continue to collect biomarker data for an extended period of time before making a decision based on the survival prediction. For example, in the ICU, healthcare professionals measure biomarkers such as blood pressure, oxygen level, and temperature to monitor the patients and predict their risks, but they must make a decision before it is too late to take action and operate on the patient. Section \ref{sec:simulationsetup} defines the parameter values and the methods to generate longitudinal and time-to-event data for this simulation study, and the results are presented in Section \ref{sec:simulationresults}.

\subsection{Setup} \label{sec:simulationsetup}
We consider two cases for the simulation study, coded in R version 4.4.0 \citep{R}. We first outline the steps to generate longitudinal and time-to-event data for fitting the joint model expressed in Equation \eqref{jm}. For both scenarios, we use the following structure of the joint model:
\begin{equation*} \label{jmsim}
h_i(t | \textbf{M}_i(t)) = h_0(t) \exp\left\{\gamma_1 w_{1i} + \gamma_2 w_{2i} + \alpha \textbf{m}_i(t)  \right\},
\end{equation*}
where 
\begin{equation*} \label{jmbiomarker}
\mathbf{m}_i(t) = \beta_0 + b_{0i} + (\beta_1 + b_{1i}) \times time
\end{equation*}
and $w_{1i}$ and $w_{2i}$ are each non-time-varying covariates associated with the hazard.

To generate longitudinal data for $ n = 2000$ in the simulation study, we use the following equation: 
\begin{equation*} \label{longmodel}
Y_{ij} = \beta_0 + b_{0i} + (\beta_1 + b_{1i}) \times time + \boldsymbol{\epsilon}_{i},
\end{equation*}
where $\mathbf{b}_i \sim MVN(\mathbf{0}, \mathbf{D})$ such that $\mathbf{D} = $ \( \begin{pmatrix} \tau_0^2 & \tau_0 \tau_1 \tau_{01}  \\ \tau_0 \tau_1 \tau_{01} & \tau_1^2 \end{pmatrix} \); $\boldsymbol{\epsilon}_{i} $ is the random error associated with individual $i$ such that $\boldsymbol{\epsilon}_{i} \sim MVN(\mathbf{0}, \sigma^2 \mathbf{I}_{n_i})$; and $time$ is a sequence of variables from 0 to 10 increasing by 0.5.  The values used to generate the longitudinal data are identified in Table \ref{tab:longparameter}.

Using these longitudinal measurements, we generate the event time for each individual. We generate survival times to simulate Cox proportional hazard models with continuous time-varying covariates, assuming that the survival times follow a Weibull distribution \cite{austin2012generating}. Survival times for the participants in our simulation study are generated using Equation \ref{GenT}, where $w_1 \sim Unif(-1.73, 1.73)$, $w_2 \sim N(0, 0.7)$. 

\begin{equation} \label{GenT}
    T = \left[ \log \left\{\frac{-\log(u) (1+v)}{\exp(\gamma_1 w_1 + \gamma_2 w_2 + \alpha(\beta_0 + b_0)) \lambda v \alpha }  +1  \right\} \frac{1}{\beta_1 + b_1} \right]^{1/(1+v)}
\end{equation}

As mentioned previously, we consider two simulation scenarios. The parameter values used to generate the two cases are identified in Table \ref{tab:alphas}. Notice that in Scenario 1, $\alpha$ is much greater than $\gamma_1$ and $\gamma_2$ compared to that in Scenario 2; the association between the biomarker and the event time is much greater than the association between the baseline covariates and the event time in Scenario 1. In Scenario 2, the magnitude of association between the baseline covariates and the biomarker with the event time is more similar. Finally, we consider censoring. In this simulation study, we choose right random censoring, where the censoring time of individual $i$ is generated as $C_i \sim Unif(1, C)$. $C$ is $7$ in Scenario 1, and $4$ in Scenario 2. 

\begin{table}[!h]
\caption{Parameter values used to generate the longitudinal data for each of the 2000 individuals in our simulation study.\label{tab:longparameter}}
\centering
\begin{tabular}{lc}
\toprule
\textbf{Parameter} & \textbf{Value} \\
\midrule
$\beta_0$ & -1.35 \\
$\beta_1$ & 0.3 \\
$\tau_0$ & 0.27 \\
$\tau_1$ & 0.08 \\
$\tau_{01}$ & 0.2 \\
$\sigma$ & 0.25 \\
\bottomrule
\end{tabular}
\end{table}

\begin{center}
\begin{table*}[!h]
\begin{threeparttable}
\caption{Parameter values used to generate the event time for each of the 2000 individuals in the two simulation study scenarios.\label{tab:alphas}}
\bigskip
\begin{tabular*}{\textwidth}{@{\extracolsep\fill}llll@{}}
\toprule
\multicolumn{2}{@{}l}{\textbf{Scenario 1$^{\tnote{\bf a}}$}} & \multicolumn{2}{@{}l}{\textbf{Scenario 2$^{\tnote{\bf b}}$}} \\\cmidrule{1-2}\cmidrule{3-4}
\textbf{Parameter} & \textbf{Value} & \textbf{Parameter} & \textbf{Value} \\
\midrule
$\lambda$ & 0.5 & $\lambda$ & 0.5 \\
$v$ & 1.03 & $v$ & 1.03 \\
$\alpha$ & 4.5 &  $\alpha$ & 4.5  \\
$\gamma_1$ & 0.5 & $\gamma_1$ & 3.5 \\
$\gamma_2$ & 1.5 & $\gamma_2$ & 3.5 \\
\bottomrule
\end{tabular*}
\begin{tablenotes}
\item[$^{\rm a}$] The coefficient of the biomarker to the event-time response variable is much greater compared to the coefficients of the fixed covariates.
\item[$^{\rm b}$] The coefficient of the biomarker to the event-time response variable is similar in magnitude to the coefficients of the fixed covariates.
\end{tablenotes}
\end{threeparttable}
\end{table*}
\end{center}

\subsection{Results} \label{sec:simulationresults}
In this simulation study, we generate five datasets for the two scenarios. Table \ref{tab:simulationresults} shows the average estimated Brier score for different subpopulation proportions across the five datasets. As explained in Section \ref{sec:algorithm}, we use repeated cross-validation to determine the optimal $M_p$, i.e. the proportion of the training data for fitting a personalized joint model. We implement a 5-fold CV repeated ten times, meaning that we can also find the standard error and obtain an approximate $100 \times (1-\alpha)\%$ confidence interval around the estimate of the Brier score for each subpopulation proportion using the following equation \citep{bates2023cross, hastie2009elements}:
\begin{equation*} \label{CI}
    \left (\bar{e} - z_{1 - \alpha/2} \widehat{SE}, \bar{e} + z_{1 - \alpha/2} \widehat{SE} \right),
\end{equation*}
where 
\begin{equation*} \label{SE}
    \widehat{SE} = \frac{1}{\sqrt{n}} \times \sqrt{\frac{1}{n-1} \sum^n_{i=1}(e_i - \bar{e})^2}
\end{equation*}
such that $n$ is the number of folds (50 in this scenario), $e_i$ is the estimate from the $i$th fold, $\bar{e}$ is the average of $e_i$ across all folds, and $z_{1 - \alpha/2}$ is the $ 1 - \alpha/2$ quantile of the standard normal distribution. This approximation should be reasonable assuming the Brier score is not too close to 0.

\begin{figure}[ht]
    \centering
    \includegraphics[width=1\linewidth]{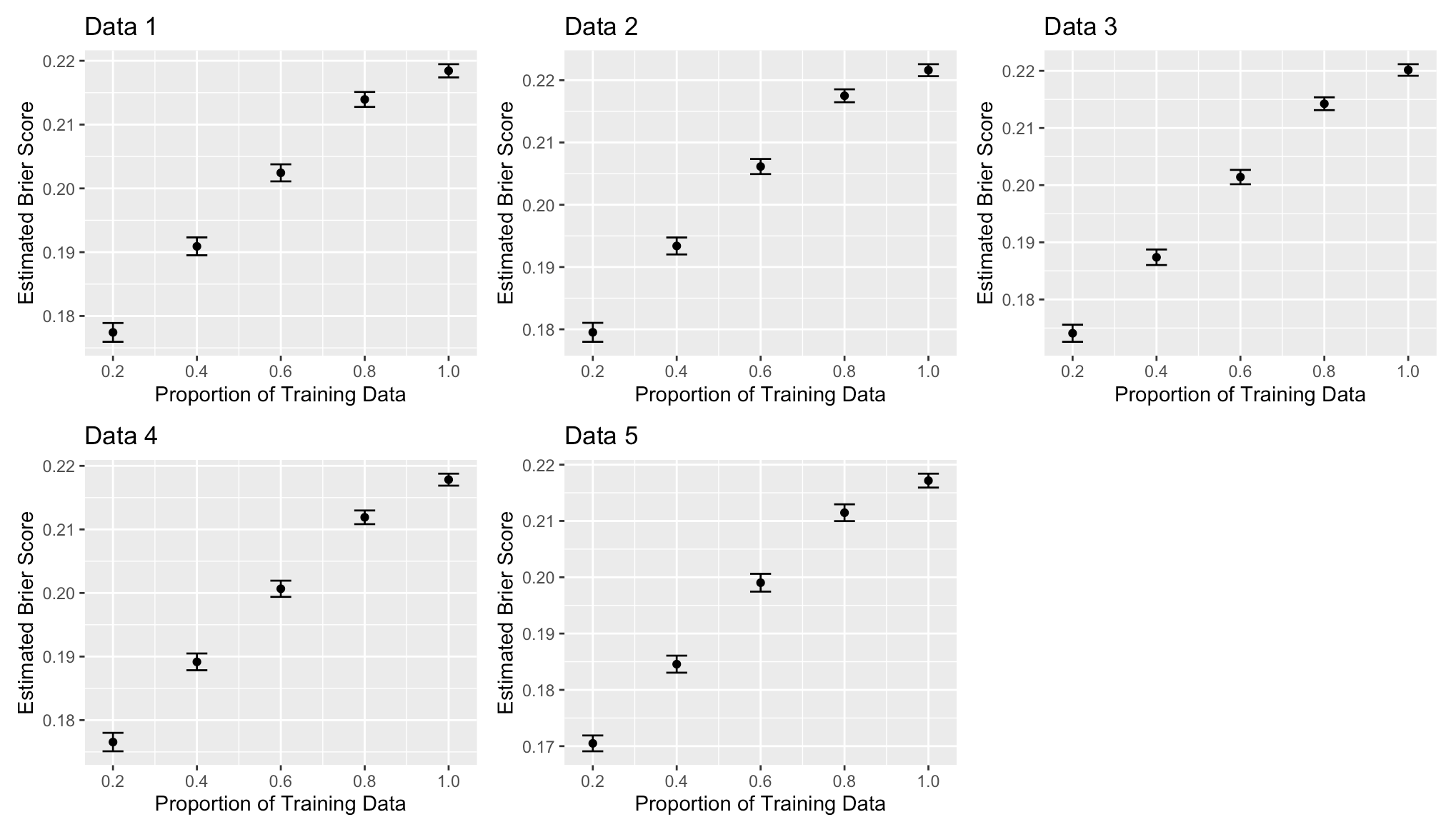}
    \caption{The estimated Brier score of the five simulated datasets for Scenario 1, with an approximate $95\%$ confidence interval around each estimate. We can see that 0.2 results in the smallest estimated Brier score across all datasets for the five values considered for the proportion of training data.}
    \label{fig:Scenario1}
\end{figure}

\begin{figure}[ht]
    \centering
    \includegraphics[width=1\linewidth]{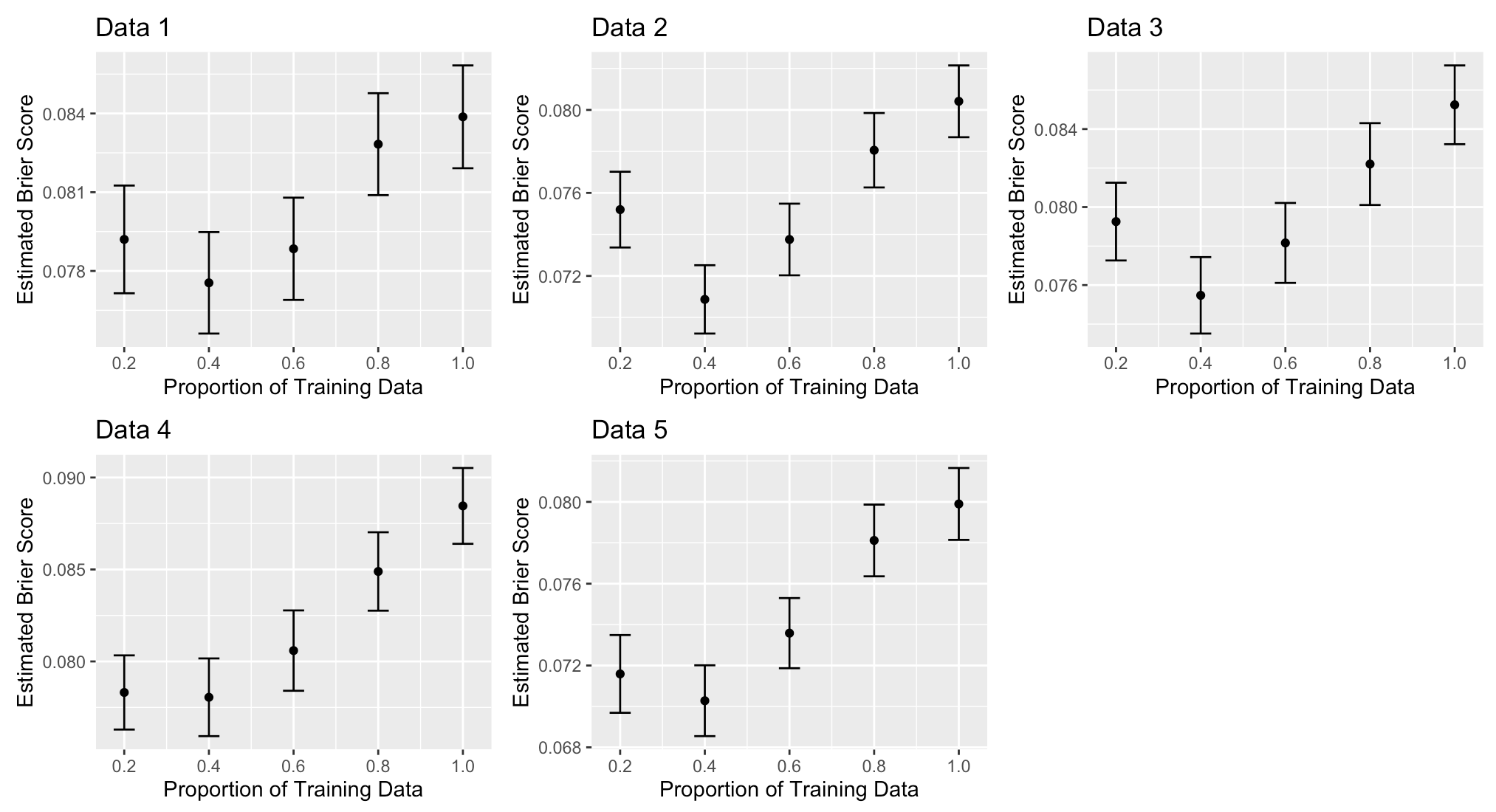}
    \caption{The estimated Brier score of the five simulated datasets for Scenario 2, with an approximate $95\%$ confidence interval around each estimate. We can see that 0.4 results in the smallest estimated Brier score across all datasets for the five values considered for the proportion of training data.}
    \label{fig:Scenario2}
\end{figure}

Recall that a smaller Brier score indicates a better predictive performance. Figure \ref{fig:Scenario1} and Figure \ref{fig:Scenario2} show the results from the five simulated datasets for each scenario with confidence intervals around the estimate for each subpopulation proportion. For Scenario 1, we can see that the estimated Brier score increases as the subpopulation proportion increases; for Scenario 2, the estimated Brier score is the lowest for $M_p = 0.4$ across the five values considered for $M_p$. For Scenario 1, where the association between the biomarker and the response is greater in comparison to the association between the non-time-varying covariates and the response, the average estimated Brier score is minimized when $M_p$ is 0.2 - that is, fitting the joint model on $20\%$ of the training data that is the most similar increases the predictive performance of the model than using the entire training data. For Scenario 2, where the association between the biomarker and the response is similar in size to that of non-time-varying covariates and the response, the average estimated Brier score is minimized when $M_p$ is 0.4. When we fit a model on targeted training data, specifically $40\%$ of the training data, we can improve the predictive performance of the model,  especially when compared to using a much larger percentage of similar individuals in the training data.  

In conclusion, this simulation study has shown that by training the model on only the individuals in the training data that are the most similar (on a set of a priori chosen covariates) to the individual that we are predicting, we can improve the predictive performance of the model, as shown by the estimated Brier score. The joint model has better prediction as the dataset becomes more targeted and there is less noise in the data for making a targeted prediction. However, the subpopulation proportion that minimizes the estimated Brier score varies depending on the dataset. For Scenario 1, it is shown that $20\%$ of the original training data is the best subpopulation size among the proportion of training set size considered; for Scenario 2, it is $40\%$ of the original training data. Note that the estimated Brier score of Scenario 2 is lower than that of Scenario 1 across all five values considered for the subpopulation proportion. This is because in Scenario 2, we use longitudinal information available up to time $t=1$ and make a prediction for time $u = 3$, whereas in Scenario 1, we use longitudinal information available up to time $t=1$ and make a prediction for a time period further into the future, which is $u=4$. Therefore, as expected, the accuracy of the model is better for $u=3$, or Scenario 2.

\clearpage
\begin{center}
\begin{table*}[!h]
\caption{A table showing the average of the estimated dynamic prediction error of joint models for different subpopulation sizes across five simulation runs for each simulation scenario, displayed in Figures \ref{fig:Scenario1} and \ref{fig:Scenario2}. The bolded value indicates the smallest estimated Brier score in each simulation case. \label{tab:simulationresults}}
\begin{tabular*}{\textwidth}{@{\extracolsep\fill}llll@{}}
\toprule
\multicolumn{2}{@{}l}{\textbf{Scenario 1}} & \multicolumn{2}{@{}l}{\textbf{Scenario 2}} \\\cmidrule{1-2}\cmidrule{3-4}
\textbf{$M_p$} & \textbf{Estimated Brier score} & \textbf{$M_p$} & \textbf{Estimated Brier score} \\
\midrule
\textbf{0.2} & \textbf{0.1756} & 0.2 & 0.0767 \\
0.4 & 0.1891 & \textbf{0.4} & \textbf{0.0744} \\
0.6 & 0.2019 & 0.6 & 0.0770  \\
0.8 & 0.2138 & 0.8 & 0.0812 \\
1.0 & 0.2190 & 1.0 & 0.0836 \\
\bottomrule
\end{tabular*}
\end{table*}
\end{center}

\section{Data Analysis}\label{sec:dataanalysis}

\begin{figure}[ht]
    \centering
    \includegraphics[width=0.6\linewidth]{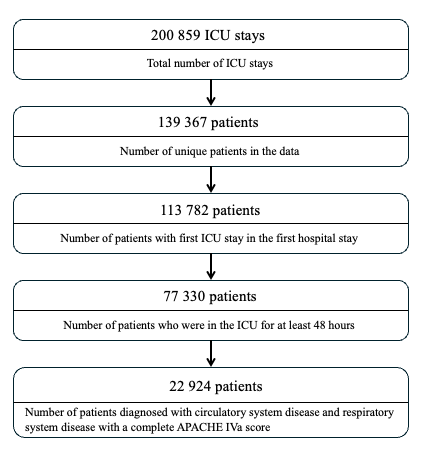}
    \caption{Flow chart of cohort selection from the eICU database for data analysis.}
    \label{fig:flowchart}
\end{figure}

We apply our proposed algorithm in Section \ref{sec:algorithm} to the dataset available in the eICU Collaborative Research Database, which covers patients admitted to critical care units in 2014 and 2015 throughout the continental United States, containing 200 859 unit encounters for 139 367 unique patients. Noting some patients have more than one hospital stay in the database, or even multiple ICU stays within a single hospital stay, we focus only on the first hospital stay of a given patient, and the first unit (ICU) stay within that first hospital stay, in order to get an independent dataset with each patient having only one record. This narrows down the number of patients to 113 782. From this dataset, we consider a subset of 77 330 patients who were in the ICU for at least 48 hours so that we have enough longitudinal information to make a prediction of the survival probability given biomarker measurements for the next 24 hours following their first day in the ICU. We require the first 24 hours in the ICU in order to allow the severity of illness score, APACHE IVa \citep{zimmerman2006acute} to be calculated, which is a commonly used prediction tool in critical care medicine. In our data analysis, we consider the patients with complete APACHE IVa score diagnosed with circulatory system disease and respiratory system disease, which contains 22 924 patients. This process is illustrated in Figure \ref{fig:flowchart}.

We randomly select 3000 patients from this dataset for computational efficiency reasons to demonstrate that we can improve the dynamic prediction of joint longitudinal-survival models using a similarity-based approach. We use the following features to fit a personalized joint model: 
\begin{itemize}
    \item Demographic data of the patient: age, gender, weight, height
    \item Patient status: the total number of laboratory result records collected during the first 24 hours after ICU admission; unit discharge status, signifying whether the patient left the unit alive or expired; unit discharge offset, or the number of minutes from unit admit time that the patient was discharged from the unit; APACHE IVa score, a predictor for mortality that incorporates patient information collected within the first 24 hours of the ICU stay, including both demographic information about the patient and physiologic data \citep{zimmerman2006acute}.
    \item Vital signs measurements: SpO2 and respiration; note that these values are available every five minutes, representing five-minute median values in the eICU database, but we take the average of each of the vital sign measurements every hour to obtain a smoother collection of these vital signs for each patient.
\end{itemize}

Then, the following joint model is fit using the dataset:
\begin{align*} \label{jm data analysis}
h_i(t | \textbf{M}_i(t)) &= h_0(t) \exp\bigg\{\gamma_1 w_{1i} + \gamma_2 w_{2i} + \gamma_3 w_{3i} + \gamma_4 w_{4i} + \gamma_5 w_{5i} + \gamma_6 w_{6i} \nonumber\\
&+ \gamma_7 w_{7i} + \gamma_8 w_{8i} + \alpha_1 \textbf{m}_{1i}(t) + \alpha_2 \textbf{m}_{2i}(t)  \bigg\},
\end{align*}
where $\textbf{m}_{1i}(t)$ is the standardized measurement of SpO2 at time $t$ such that
\begin{equation*} \label{spo2}
\begin{aligned}
\mathbf{m}_{1i}(t)
={}& \beta_{0(1)} + b_{0i(1)}
+ (\beta_{1(1)} + b_{1i(1)}) \times time \\
&+ \beta_{2(1)} age
+ \beta_{3(1)} weight
+ \beta_{4(1)} height
+ \beta_{5(1)} lab\_count.
\end{aligned}
\end{equation*}
and $\textbf{m}_{2i}(t)$ is the standardized measurement of heartrate at time $t$ such that

\begin{equation*} \label{respiration}
\begin{aligned}
\mathbf{m}_{2i}(t)
={}& \beta_{0(2)} + b_{0i(2)}
+ (\beta_{1(2)} + b_{1i(2)}) \times time \\
&+ \beta_{2(2)} age
+ \beta_{3(2)} weight
+ \beta_{4(2)} lab\_count
+ \beta_{5(2)} gender.
\end{aligned}
\end{equation*}

\noindent and $w_{1i}$, $w_{2i}$, $w_{3i}$, and $w_{4i}$ signify gender, age (standardized), APACHE IVa score (standardized), and weight of (standardized) patient $i$ respectively, with the response variable being the number of minutes from the first 24 hours in the ICU that the patient was discharged from the unit. $w_{5i}$ and $w_{6i}$ represent the two functional principal components that explain at least 90\% of the variation in the data for SpO2 and $w_{7i}$ and $w_{8i}$ represent the two functional principal components that explain at least 90\% of the variation in the data for respiration. The variables included in the final models were selected based on the Deviance Information Criterion (DIC) and the Watanabe-Akaike Information Criterion (WAIC). Models with lower values of both criteria were preferred, as lower values indicate better expected model performance after accounting for model complexity.

To measure the similarity between the patients for fitting a personalized predictive model, we use all of the following predictors, which can be categorized into two groups. The first set of predictors is the baseline covariates and the information we gather from the first 24 hours of being admitted into the ICU. These include each patient's standardized age, gender, weight, height, lab count, standardized APACHE IVa score, the first three functional principal components and the maximum, minimum, and median of SpO2 and respiration in the first 24 hours. The second set of predictors is the information we gather from the biomarker measurements in the next 24 hours since being admitted to the ICU for one day (i.e., hours 24+ to 48): the three functional principal components for each of the vital sign measurements that explain at least 95\% of the variation in the data for the next 24 hours after the first day since admitted to the ICU. Note that, although two functional principal components per vital sign were retained in the joint model based on model selection using DIC and WAIC, three FPCs per vital sign are used here to calculate similarity, as they collectively explain at least 95\% of the variation in the data. We use the biomarker measurements for 24 hours after the first day of being admitted to the ICU only for similarity calculation because we want to make a prediction for the survival rate at some future time $u$ given biomarker information up to $t$, which is 24 hours in this data analysis. We consider $20\%, 40\%, 60\%, 80\%$, and $100\%$ of the population of the training data as the subpopulation size to compare the estimated Brier score of the joint model for predicting the probability of survival at 5000 minutes (approximately 83 hours or 3.5 days) after the first 24 hours in the ICU, given longitudinal measurements from hour 24+ to 48 in the ICU.

\subsection{Results} \label{sec:dataresults}

\begin{figure}[ht]
    \centering
    \includegraphics[width=0.9\linewidth]{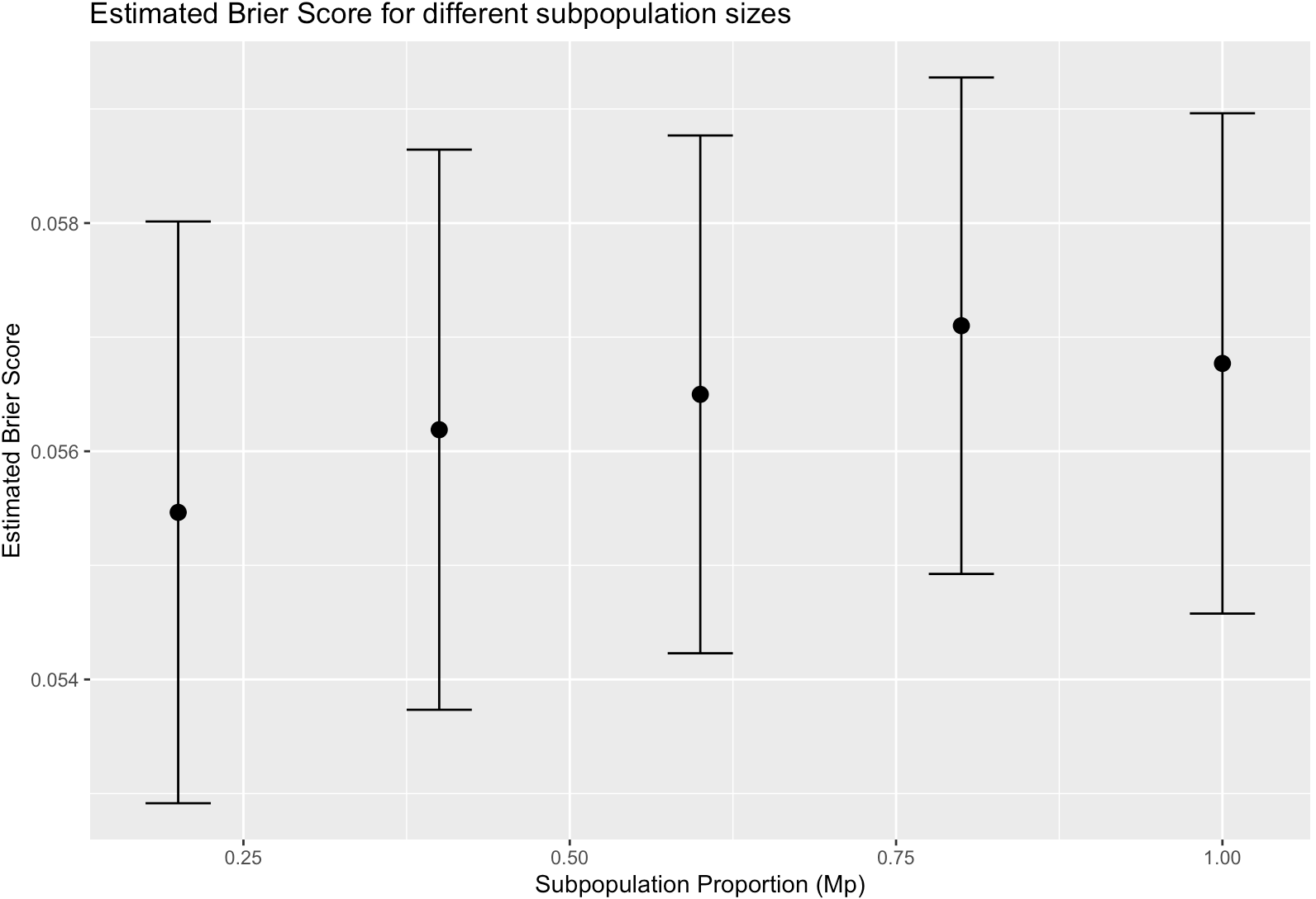}
    \caption{The estimated Brier score of the eICU dataset across five different subpopulation proportions. We can see that 0.2 results in the smallest estimated Brier score.}
    \label{fig:eicu}
\end{figure}

\begin{table}[!h]
\centering
\caption{The estimated Brier score for five different subpopulation proportions for the eICU data. When the subpopulation proportion is 0.2, the estimated Brier score is minimized. The columns ``Lower" and ``Upper" represent the 95\% confidence interval of the estimates.}
\label{tab:eicu}
\begin{tabular}{cccc}
\toprule
\textbf{$M_p$} & \textbf{Estimated BrS} & \textbf{Lower} & \textbf{Upper}\\
\midrule
\textbf{0.2} & \textbf{0.0555} & 0.0529 & 0.0580\\
0.4 & 0.0562 & 0.0537 & 0.0586 \\
0.6 & 0.0565 & 0.0542 & 0.0588\\
0.8 & 0.0571 & 0.0549 & 0.0593\\
1.0 & 0.0568 & 0.0546 & 0.0590\\
\bottomrule
\end{tabular}
\end{table}

From Figure \ref{fig:eicu} and Table \ref{tab:eicu}, we can see that the model performs the best in estimating the survival probability at time 5000 minutes with longitudinal measurements until 1440 minutes (i.e., 24 hours) after the first day in the ICU when the proportion of the data is 20\% of the training data - the estimated Brier score is approximately 0.0555. Note that these estimated values are not comparable to the results from the simulation study - not only is the model from the data analysis more complex, but the prevalence in the dataset also differs. In the simulation study, around 48\% of the patients in the data experienced the event, whereas only about 11\% of the patients experienced the event in the data analysis. While the pattern of the estimated Brier score plotted against $M_p$ is different between the simulation study and the data analysis, the conclusion is the same in that the predictive performance of joint models with longitudinal and time-to-event data can be improved by fitting the model on a more personalized data - 20\% and 40\% being the optimal subpopulation proportions from the two cases in the simulation study, and 20\% for the data analysis, though certainly the improvement with the subpopulation approach (compared to using the entire training data) is not as favourable as seen in the two simulation results.

In the validation step, we evaluate the predictive performance of the joint model with personalized predictive modeling for the proportion of training data equal to $20$\%, which was shown to have the most optimal outcome from the data analysis. We randomly sample 3000 patients from the hold-out validation sample and apply the algorithm outlined in Section \ref{sec:algorithm}, with the proportion $M_p$ set to 20\%. When sampling the data from the hold-out validation set, we used stratified sampling so that the event rate is similar to that of the training-testing dataset; the resulting estimated Brier score was 0.0548 for $M_p = 0.2$ compared to 0.0570 for $M_p = 1.0$.

\section{Discussion}\label{sec:discussion}

In this work, we have shown how we can improve the dynamic predictive performance of joint models for longitudinal and time-to-event data using our proposed similarity-based modeling approach. Joint models are commonly used in follow-up studies and have three notable features that make them quite attractive for model predictions: first, they use the entire longitudinal information available up to a time point instead of using only the most recent measurement; second, they allow for dynamic predictions by updating the survival probability as more longitudinal measurements become available; third, they allow for individualized predictions by incorporating random effects into the model. Our goal is to improve the predictive accuracy of joint models by personalizing the training data to the individual we are making the predictions for, using our proposed method outlined in Section \ref{sec:algorithm}. Note that the cross-validation process to identify the optimal $M_p$ was a complex extension of the process outlined in Lee and Dubin \citep{lee2015personalized}, but applying the similarity-based approach to train the joint model on personalized data to improve the dynamic prediction of joint models is the novelty introduced here. 

More specifically, we use patient similarity to identify the individuals in the training data that are the most similar to the individual that we want to predict for, and train the model on $M_p \times 100\%$ of the most similar training data. The optimal size of this personalized subset of the data varied depending on the simulation scenario, but ultimately, we can decrease the estimated Brier score and thus improve the predictive performance of the joint models by training the model on a subpopulation of the dataset. Our simulation study with repeated 5-fold cross-validation showed that 20\% was the best subpopulation proportion among the considered values for Scenario 1, and 40\% was the best subpopulation proportion among the considered values for Scenario 2. In other words, our research can indicate that some advancements and improvements can be seen in the dynamic predictive performance of joint models by taking a more targeted sample from the training data - this agrees with past research in patient similarity literature, particularly when modeling binary response data (e.g. \cite{lee2015personalized, krikella2025personalized}).

Note that Scenario 1 in the simulation study shows that as we decrease the proportion of the training data that we are fitting the joint model - that is, as our dataset becomes more targeted with less noise - the predictive performance continues to improve. We do not show the results for proportions smaller than $M_p = 0.2$ because it leads to optimization errors when fitting a linear mixed-effects model. In our simulation study, we train the model on a personalized dataset of size $m = 360$ to estimate the Brier score when $M_p = 0.2$. However, when this value is smaller, it leads to optimization errors for some datasets. In other words, the optimal value of $20\%$ is specific to the dataset simulated in our study - for a different dataset with more samples, for example, the optimal $M_p$ might be less than $20\%$ of the original training data since a smaller proportion of the data may not lead to optimization errors if the original training data size is sufficiently large.

In the data analysis section, we apply our algorithm to a real dataset, specifically the eICU database, to demonstrate how our method can be useful in real-life scenarios by effectively increasing the predictive performance of joint models on patients in intensive care units. The eICU database offers rich information in both the patients' demographic information and lab and vital sign results during the ICU stay - we increase the complexity of the joint model from the joint models in the simulation study by incorporating more covariates. The point estimate of the Brier score suggests that it is minimized when $M_p = 0.2$ compared to $M_p = 0.1$, and thus we can improve the predictive performance of joint models by training the model on a targeted subpopulation of the training data. However, the magnitude of this improvement is considerably smaller than what was observed in our two simulation scenarios. We suspect that this may be due to the correlation structure among the predictors, which are not present in the simulated datasets, or a different functional form of the longitudinal predictors must be specified in the joint model (such as incorporating the slope of the time-dependent features, instead of just using the current value of the underlying longitudinal trajectory).

There are several areas of future work in this study. First, as mentioned above, our results show more favourability in terms of model predictive performance in the simulation results than in the data analysis, in that there is a greater improvement when the joint model is trained on personalized data for the simulation study compared to the data analysis; thus, further investigation is necessary. Perhaps a different subset of patients in the eICU dataset, or a different ICU database altogether, may show results that more closely align with the simulation study. It would also be worthwhile to investigate the correlation structure among predictors and the functional form of the longitudinal predictors in the joint model for the data analysis and evaluate how these factors influence the effectiveness of the similarity-based approach.  Second, we only consider one measure of predictive accuracy, namely the Brier score. We hope to evaluate the predictive performance of the models using additional metrics, such as log loss and calibration. In addition, we will consider the scaled Brier score to account for outcome prevalence, which allows for more meaningful comparisons across different datasets. Third, we have considered very specific values of $t$ and $u$ in our simulation study. The choices for $t$ and $u$ are context-dependent. For example, relatively small values of $t$ and $u$ are used with the eICU database, as clinical decisions must be made quickly before patient deterioration. In contrast, other applications, such as monitoring cancer biomarkers to predict recurrence, may involve much longer time horizons, ranging from months to years. Therefore, future work will extend the simulation study to incorporate a wider range of $t$ and $u$ values in order to assess whether these choices influence tuning for the optimal $M_p$. Furthermore, we only work with the cosine similarity metric to quantify the similarity between the individuals in the training data and the testing data - it may be worthwhile to study how the predictive accuracy of joint models changes under the similarity setting with different approaches to measure similarity, such as Mahalanobis distance or clustering methods. Finally, we hope to study the performance of joint models with the introduction of missing data in both biomarkers and time-invariant covariates. Currently, in our simulation study, we assume that all measurements are available - we would like to study how our results change with missing data, both for the type of missing and when we use different approaches to deal with missing data, such as different imputation methods. 

Finally, as mentioned earlier, a big challenge of our work was the computational burden. We have taken measures to increase the computational efficiency, such as fitting the Bayesian model coded in C++ and parallelizing our algorithm. However, since we fit a personalized model for each individual in the testing data for every fold in the repeated 5-fold cross-validation, we fit thousands of models to tune the optimal subpopulation size - this increases the computational time dramatically. An approach for improving computational efficiency is to leverage recently developed software such as the R package \textit{pencal}, which implements a non-Bayesian approach for dynamic prediction \citep{Signorelli2024pencal}. Investigating whether \textit{pencal} can be incorporated into or adapted for our personalized modeling framework represents a potential direction for future research.

\bibliographystyle{References}  
\bibliography{References}     

\end{document}